\documentclass[showpacs,amsmath,amssymb,11pt]{article}
\usepackage{graphicx,color}
\usepackage{amsmath}
\usepackage{amssymb}
\usepackage{amsmath,amsfonts,latexsym,graphicx,amssymb}
\usepackage{amsfonts}
\usepackage{amssymb}
\usepackage{mathrsfs}
\usepackage{amsthm}
\usepackage[utf8]{inputenc}
\usepackage{graphicx}
\usepackage{amsmath}
\newcommand{\ot}{{\,\otimes\,}}
\newcommand{{\Cd}}{{\mathbb{C}^d}}
\newcommand{{\Rn}}{{\mathbb{R}^n}}

\def\oper{{\mathchoice{\rm 1\mskip-4mu l}{\rm 1\mskip-4mu l}%
{\rm 1\mskip-4.5mu l}{\rm 1\mskip-5mu l}}}
\def\<{\langle}
\def\>{\rangle}

\newtheorem{cor}{Corollary}

\newtheorem{ex}{Example}

\newtheorem{pro}{Proposition}
\newtheorem{remark}{Remark}

\date{}

\begin{document}
\title{\textbf{
On pseudo-stochastic matrices \\ and pseudo-positive maps}}
\author{D. Chru\'sci\'nski$^1$\footnote{email: darch@fizyka.umk.pl}, V.I. Man'ko$^{2,3}$, G. Marmo$^{4,5}$, and F. Ventriglia$^{4,5}$ \\
 \\
$^1$Institute of Physics, Faculty of Physics, Astronomy and Informatics, \\
Nicolaus Copernicus University, Grudzi{a}dzka 5, 87--100 Toru\'n, Poland \\ \\
$^2$P. N. Lebedev Physical Institute, Russian Academy of Sciences, 119991 Moscow,
Russia \\ \\
$^3$Moscow Institute of Physics and Technology, Dolgoprudni, Moscow Region,
Russia \\ \\
$^4$Dipartimento di Fisica and MECENAS, \\ Universit\`a di Napoli “Federico II”,
I-80126 Napoli, Italy \\ \\
$^5$INFN, Sezione di Napoli, I-80126 Napoli, Italy}

\maketitle

\begin{abstract}

Stochastic matrices and positive maps in matrix algebras proved to be very important tools for analysing classical and quantum systems. In particular they represent a natural set of transformations for classical and quantum states, respectively.
Here we introduce the notion of pseudo-stochastic matrices and consider their
semigroup property. Unlike stochastic matrices, pseudo-stochastic matrices are permitted to have matrix elements
which are negative while respecting the requirement that the sum of the
elements of each column is one. They also allow for convex combinations, and
carry a Lie group structure which permits the introduction of Lie algebra
generators. The quantum analog of a pseudo-stochastic matrix exists and is
called a pseudo-positive map. They have the property of transforming a subset
of quantum states (characterized by maximal purity or minimal von Neumann
entropy requirements) into quantum states. Examples of qubit dynamics
connected with ``diamond" sets of stochastic matrices and
pseudo-positive maps are dealt with.

\end{abstract}

\section{Introduction}

Stochastic matrices and linear positive maps are well established tools for dealing with many problems in stochastic processes, stochastic evolution and quantum information theory \cite{Kampen,Gardiner,Paulsen}.  A stochastic matrix $T_{ij}$ satisfies two basic properties: $T_{ij} \geq 0$ and $\sum_i T_{ij} =1 $.
These properties guarantee that stochastic matrices map probability vector into a probability vector and hence may be used to describe legitimate operations on the classical states (described by probability vectors).


States of quantum systems are described by density matrices, their
evolution is usually described by positive or completely positive
maps \cite{Stinespring,Rao,Choi,Kraus}. The evolution equation of Markovian type for density matrices was
introduced by Kossakowski \cite{Kos} and further elaborated by Gorini, Kossakowski, Sudarshan \cite{GKS} and independently by Lindblad \cite{Lindblad}. Nowadays, open quantum systems and their
dynamical features are attracting increasing attention \cite{Alicki,Breuer,Weiss,RIVAS}.
They are of paramount importance in the study of the
interaction between a quantum system and its environment,
causing dissipation, decay, and decoherence \cite{Schlos}.


It was observed \cite{Mancini,Olga,Ibort} that quantum states may also be described
by tomographic probability distributions both for finite (qudit) and
infinite (photon quadratures) dimensional Hilbert spaces. According to this
picture standard quantum evolution can be related with the evolution of
probabilities describing the quantum states, it was first observed on
simple examples \cite{Filipov,Trask} and then considered in its
generality \cite{PS-DBFV} that the evolution of probability vectors is related
with the analog of stochastic matrices which can have negative matrix
elements. The violation of positivity is associated with the observation
that probability vectors describing quantum states occupy only a subset of
the simplex. Such a phenomenon does not seem to be well known in the
existing literature.
It is worthy to mention that recently \cite{Holevo} non-positive  maps of
Gaussian states have made their appearance in the discussion of properties
of quantum channels. A non-positive map obtained by rescalling the argument of the Wigner function was used also in \cite{NON}.

In this paper we would like to study linear maps on the space of
probability vectors which need not be stochastic but are only
pseudo-stochastic, as we are going to call them. These maps naturally appear
when we only consider the transformation of subdomains in the
simplex. Another aspect of this paper is the introduction of non-positive
maps as maps acting on the space of density matrices.
It is natural to consider the relation between non-positive maps of density
states with the notion of pseudo-stochastic matrices.
The subset of density matrices associated with
subdomains of the probability vectors are just the objects which can be
transformed by means of pseudo-positive maps.
To this aim, we consider subsets of density matrices and characterize them
by maximal purity or minimal entropy requirements. To illustrate
these ideas we reconsider in details the dynamics of qubit states
following \cite{PLA,PRA}. Pseudo-positive maps which are positive only on the convex subset $K$ may be considered as witnesses of ``not being an element of $K$" in the same way as positive but non completely positive maps are witnesses of being non-separable state \cite{HHHH}.

In this paper we point out the importance of the clear understanding of the introduction of the notion of pseudo-stochastic matrices and pseudo-positive maps for quantum information. It turns out that the use of these matrices and maps is natural in quantum mechanics and represents another aspects of classical-to-quantum transition. In the classical setting it was sufficient to use stochastic matrices for the description of kinetic phenomena associated with random variables and probability distributions as well as with their time evolution. In the quantum setting the evolution of states considered in the framework of probability distributions demands the use of pseudo-stochastic matrices. Moreover, when considering the density matrices and their evolution as well as quantum channels in all their diverse facets we need to introduce pseudo-positive maps. It appears that these maps are new elements to be taken into account for  the analysis of quantum correlation properties and the analysis of quantum information processes.

The paper is organized as follows.
In section 2 we study the semigroup of stochastic and pseudo-stochastic
matrices and identify convex subsets of these matrices. In section 3 we provide an instructive example of such subsets which we call ``diamond"  subsets.  In section 4 we discuss an example of classical evolution of a 2-level system in connection to "diamond" subsets
in $\mathbb{R}^2$.  In section 5 the quantum pseudo-positive maps are dealt with and
considered on the example of qubit state dynamics.In section 6 we draw
some conclusions and advance some perspectives . In Appendix the Lie algebra structure of the  group of pseudo-stochastic matrices for $n=2$ and $n=3$ is shortly discussed.

\section{A semigroup of pseudo-stochastic matrices}

A real $n \times n$ matrix $T$  is {\em stochastic} iff $T_{ij} \geq 0$ and  $\sum_{i=1}^n T_{ij} = 1$ \cite{Horn,Bhatia}. It defines a compact convex subset  ${\rm S}_n \subset \mathbb{R}^{n(n-1)}$. Stochastic matrices define a semigroup: if $T_1,T_2 \in {\rm S}_n$, then $T_1 T_2 \in {\rm S}_n$. It is not a group because $T$ does not need to be invertible and even if $T^{-1}$ exists it needs not belong to ${\rm S}_n$.
Actually, $T^{-1} \in {\rm S}_n$ iff $T \in {\rm Per}_n$, where ${\rm Per}_n$ denotes a set of $n \times n$ permutation matrices.
It is clear that ${\rm Per}_n$ defines a discrete group being a subgroup of the unitary group $U(n)$.
Stochastic matrices satisfying the additional condition $\sum_{j=1}^n T_{ij} = 1$ define a proper convex subset ${\rm BS}_n \subset {\rm S}_n$ of {\em bistochastic matrices}. According to the celebrated Birkhoff theorem \cite{Bhatia} any bistochastic matrix $T$ is a convex combination of permutation matrices.

Now, we relax the condition $T_{ij} \geq 0$ and call the matrix $T \in \mathbb{M}_n(\mathbb{R})$  {\em pseudo-stochastic} iff   $\sum_{i=1}^n T_{ij} = 1$. A set ${\rm PS}_n$ of $n \times n$ pseudo-stochastic matrices is isomorphic to $\mathbb{R}^{n(n-1)}$ and ${\rm S}_n$ defines a convex subset of ${\rm PS}_n$. It is clear that ${\rm PS}_n$ defines a semigroup: if $T_1,T_2 \in {\rm PS}_n$, then $T_1 T_2 \in {\rm PS}_n$. Let us observe that  if $T\in {\rm S}_n$ is invertible, then $T^{-1} \in {\rm PS}_n$.

If $T \in {\rm PS}_n$ and $T$ is invertible, then $T^{-1} \in {\rm PS}_n$. Hence
\begin{equation}\label{}
  {\rm GPS}_n = \{  T \in {\rm PS}_n \ | \  \det T \neq 0 \} \subset {\rm PS}_n\ ,
\end{equation}
defines a group of pseudo-stochastic matrices. It is a subgroup of ${\rm GL}(n,\mathbb{R})$ and contains ${\rm Per}_n$ as a discrete subgroup. Note, that ${\rm GPS}_n^+$ containing invertible matrices from ${\rm PS}_n$ such that $\det T>0$ defines a subgroup of ${\rm GPS}_n$ --- the connected component of identity.


Stochastic matrices provide mathematical representation of classical channels
\begin{equation}\label{}
  T : \mathbb{R}^n \rightarrow \mathbb{R}^n\ ,
\end{equation}
that is $T(\Sigma_n) \subset \Sigma_n$, where
\begin{equation}\label{}
  \Sigma_n = \left\{ \ \mathbf{p}=(p_1,\ldots,p_n) \in \mathbb{R}^n\ |\ p_k \geq 0\ , \sum_{k=1}^n p_k =1 \ \right\}\ ,
\end{equation}
defines a simplex of probability distributions (classical states). Consider a  convex subset $K \subset \Sigma_n$ and  define

\begin{enumerate}

\item ${\rm S}(K) \subset S_n$ such that for all $T \in {\rm S}(K)$ one has $T(K) \subset K$,

\item ${\rm PS}(K) \subset {\rm PS}_n$ such that for all $T \in {\rm PS}(K)$ one has $T(K) \subset \Sigma_n$,

\item ${\rm S}_0(K) \subset {\rm S}(K)$ such that for all $T \in {\rm S}_0(K)$ one has $T(\Sigma_n) \subset K$.

\end{enumerate}
One immediately finds
\begin{equation}\label{}
  {\rm S}_0(K) \subset {\rm S}(K) \subset S_n \subset {\rm PS}(K) \subset {\rm PS}_n \ ,
\end{equation}
and if $K = \Sigma_n$, then
\begin{equation}\label{}
  {\rm S}_0(\Sigma_n) = {\rm S}(\Sigma_n) = S_n \subset {\rm PS}(\Sigma_n) = {\rm PS}_n \ .
\end{equation}
Interestingly, if $K =\{\mathbf{p}_*\}$ with $\mathbf{p}_* = (\frac 1n,\ldots,\frac 1n)$, then  ${\rm S}(\{\mathbf{p}_*\})$ defines a set of bistochastic matrices and ${\rm S}_0(\{\mathbf{p}_*\})$ contains only one element $T_*$ defined by $(T_*)_{ij} = \frac 1n$ (a maximally mixing bistochastic matrix).

A set ${\rm S}(K)$ has a clear interpretation: a subset $K$ is $T$-invariant for all $T\in {\rm S}(K)$. Note, that if $T_1,T_2 \in {\rm S}(K)$ in general $T_1T_2$ needs not belong to ${\rm S}(K)$. However, one proves

\begin{pro} For any $T_1,T_2 \in {\rm S}_0(K)$, $T_1 T_2 \in  {\rm S}_0(K)$, that is,  ${\rm S}_0(K)$ defines a semigroup (subsemigroup of ${\rm S}_n$).
\end{pro}
%
%
Convex sets ${\rm S}_0(K)$ and ${\rm S}(K)$ contain only stochastic matrices. A set ${\rm PS}(K)$ contains also pseudo-stochastic matrices which are not stochastic. The interpretation of these matrices is provided by the following

\begin{pro} An element ${\bf p} \in \Sigma_n$ belongs to $K$ if and only if $T {\bf p} \in \Sigma_n$ for all $T \in {\rm PS}(K)$.
\end{pro}

\noindent Hence, element from ${\rm PS}(K) - S_n$ may be used to witness that ${\bf p}$ does not belong to $K$.

\begin{cor} An element ${\bf p} \in \Sigma_n$ does not belong to $K$ if and only if there exists $T  \in {\rm PS}(K) - S_n$ such that $T{\bf p} \notin \Sigma_n$.
\end{cor}

\section{Example: ``diamond" subsets}


Consider the following convex subset $K_\varepsilon$  of $\Sigma_2$
\begin{equation}\label{}
 \varepsilon \leq  p_1,p_2 \leq 1- \varepsilon \ ,
\end{equation}
with $0 \leq \varepsilon \leq \frac 12$. Clearly, $K_0 = \Sigma_2$ and $K_{\frac 12} = \{\mathbf{p}_*\}$, where $\mathbf{p}_*=(\frac 12,\frac 12)$ is the maximally mixed state. Moreover $K_{\frac 12} \subset K_{\varepsilon} \subset K_{\varepsilon'} \subset \Sigma_2$ for $\varepsilon' \leq \varepsilon$. Any $2\times 2$ pseudo-stochastic matrix may be parameterized by two real numbers $(a,b)$ as follows
\begin{equation*}
  T = \left( \begin{array}{cc} a & 1-b \\ 1-a & b \end{array} \right)\ .
\end{equation*}
Two convex sets ${\rm S}(K_\varepsilon)$ and  ${\rm PS}(K_\varepsilon)$ are represented by diamond shape bodies displayed in Figure 1: ${\rm S}(K_\varepsilon)$ corresponds to the inner violet diamond and ${\rm PS}(K_\varepsilon)$ corresponds to the outer yellow diamond. One finds for the corresponding vertices
\begin{equation}\label{}
  A = \frac{1-\varepsilon}{1-2\varepsilon}\,(1,1)\ ,\ B= \frac{\varepsilon}{1-2\varepsilon}\,(-1,-1) \ ,\ C=( \varepsilon,1-\varepsilon)\ , \ D= (1-\varepsilon,\varepsilon)\ .
\end{equation}
In this case ${\rm S}_2$ is represented by the red square $[0,1]\times [0,1]$ and ${\rm PS}_2$ is the whole $(a,b)$-plane $\mathbb{R}^2$. Finally, ${\rm S}_0(K_\varepsilon)$ corresponds to the inner dark blue square $[\varepsilon,1-\varepsilon]\times  [\varepsilon,1-\varepsilon]$. If $\varepsilon \rightarrow  \frac 12$, then ${\rm S}(K_{\frac 12})$ defines the set of bi-stochastic matrices and ${\rm PS}(K_{\frac 12})$ the set of pseudo-bistochastic matrices represented by the line $a=b$. Finally ${\rm S}_0(K_{\frac 12})$ shrinks to the point $(\frac 12,\frac 12)$. Note, that
\begin{equation}\label{}
  \det T = {\rm tr} T-1\ ,
\end{equation}
and hence $T$ is invertible iff ${\rm tr}\,T \neq 1$. It gives rise to a group of pseudo-stochastic matrices
$$  {\rm GPS}_2 = \{  T \in {\rm PS}_2 \ | \  {\rm tr}\,T \neq 1 \}\ , $$
and the proper subgroup of stochastic matrices contains only two elements
$$  {\rm GS}_2 = \{  T_0,T_1 \} \subset {\rm GPS}_2\ ,  $$
where $T_0 = \mathbb{I}_2$ and $T_1 = \sigma_x$ is a permutation matrix.
The group ${\rm GPS}_2$ has two connected components:  ${\rm GPS}_2^+$ corresponding to $\det T >0$ and ${\rm GPS}_2^-$ corresponding to $\det T <0$. ${\rm GPS}_2^+$ contains identity matrix whereas ${\rm GPS}_2^-$ contains permutation matrix. The Lie algebra properties of the "diamond" group are shortly discussed in the Appendix.


\begin{figure}[!h] \label{FIG-OPT}
\begin{center}
 \includegraphics[width=10cm]{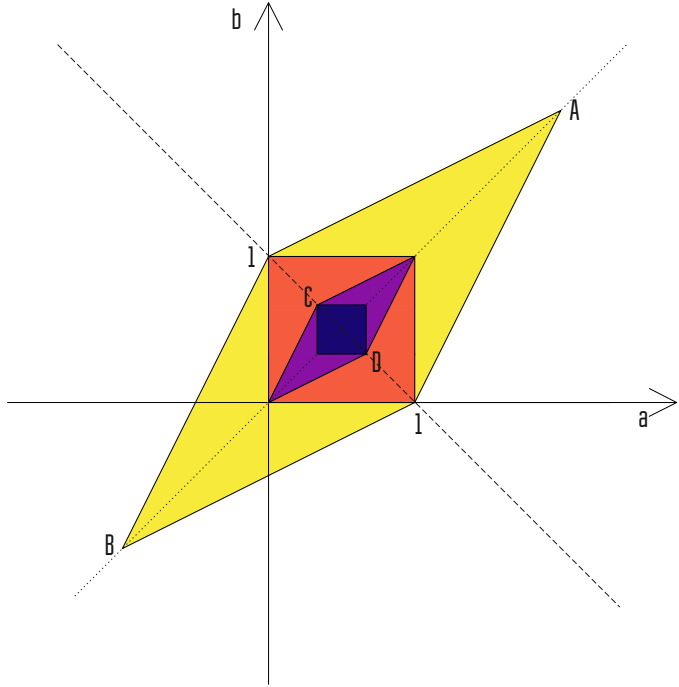}
\end{center}
\caption{[color online] Yellow diamond with vertices $A$ and $B$ corresponds to ${\rm PS}(K_\varepsilon)$ (for $\varepsilon=1/3$); red square $[0,1]\times [0,1]$ corresponds to stochastic matrices; violet diamond with vertices $C$ and $D$ corresponds to ${\rm S}(K_\varepsilon)$; dark blue square with vertices $C$ and $D$ corresponds  ${\rm S}_0(K_\varepsilon)$. A line $a=b$ passing through vertices $A$ and $B$ represents pseudo-bistochastic matrices. A line $a+b=1$ passing through vertices $C$ and $D$ represents singular pseudo-stochastic matrices. Both lines intersect in $T_*$ --- a maximally mixing bistochastic matrix.
 }
 \end{figure}




\section{Divisible dynamical maps and pseudo-stochastic propagators}

Consider now a classical evolution $\mathbf{p}_t \in \Sigma_n$ described by the dynamical map $T(t) \in {\rm S}_n$ satisfying initial condition $T(t=0) = \mathbb{I}_n$, that is, $\mathbf{p}_t = T(t) \mathbf{p}_0$. An example of such a map is provided by Markovian semi-group $T(t) = e^{tL}$, where $L \in \mathbb{M}_n(\mathbb{R})$ is the corresponding generator satisfying well known conditions \cite{Kampen}
\begin{equation}\label{}
  L_{ij} \leq 0\ , \ (i \neq j) \ ; \ \ \ \ \sum_{i=1}^n L_{ij} = 0 \ .
\end{equation}
Note, that $T(t) = e^{tL}$ is invertible $T^{-1}(t) = e^{-tL}$ and clearly $T^{-1}(t) \in {\rm PS}_n \smallsetminus {\rm S}_n$ for $t>0$.
However, the corresponding propagator
\begin{equation}\label{}
  V(t,s) := T(t) \cdot T^{-1}(s) = e^{(t-s)L}\ ,
\end{equation}
belongs to ${\rm S}_n$ for $t \geq s$. It is clear that if $L$ satisfies only $\sum_{i=1}^n L_{ij} = 0$, then $T(t) = e^{tL}$ defines a 1-parameter semigroup of pseudo-stochastic maps.

A dynamical map $T(t)$ is divisible if for any $t > s$ one has
\begin{equation}\label{TVT}
  T(t) = V(t,s) T(s)\ ,
\end{equation}
and $V(t,s) \in {\rm S}_n$. Note, that if $T(t)$ is invertible, then $V(t,s) = T(t) \cdot T^{-1}(s)$. Now, the family $T(t)$ is divisible if and only if it satisfies the time-local master equation
\begin{equation}\label{ME}
  \frac{d}{dt} T(t) = L(t) T(t)\ ,
\end{equation}
with time-dependent local generator satisfying $ L_{ij}(t) \leq 0$ for $i \neq j$, and $\sum_{i=1}^n L_{ij}(t) = 0$. The corresponding propagator is given by
\begin{equation}\label{}
  V(t,s) = T \exp\left( \int_s^t L(u) du \right)\ .
\end{equation}
General dynamical map needs not be divisible.

\begin{ex} Consider the following time-local generator for $n=2$
\begin{equation}\label{}
  L(t) =  \left(  \begin{array}{rr} -x(t) & y(t) \\ x(t) & -y(t)    \end{array} \right)\ .
\end{equation}
It is clear that it generates divisible dynamical map iff $x(t), y(t) \geq 0$ for $t \geq 0$. Note that
\begin{equation}\label{}
  L(t) =  \left(  \begin{array}{rr} y(t) & y(t) \\ x(t) & x(t)    \end{array} \right) -  \left(  \begin{array}{rr} \gamma(t) & 0 \\ 0 & \gamma(t)\end{array} \right) \ , 
\end{equation}
with $\gamma(t) = x(t) + y(t)$, and hence
\begin{equation}\label{}
  L(t)\mathbf{p} = \gamma(t) \left[ \mathbf{q}(t) (p_1+p_2) - \mathbf{p}\right]\ ,
\end{equation}
where $\mathbf{q}_t=(q_1(t),q_2(t))$, with $q_1(t) = y(t)/\gamma(t)$ and  $q_2(t) = x(t)/\gamma(t)$. Interestingly, one has
\begin{equation}\label{}
  L(t)\mathbf{q}(t) = 0 \ ,
\end{equation}
that is, $\mathbf{q}(t)$ is a time-dependent invariant vector. One easily finds the corresponding solution
\begin{equation}\label{}
 \mathbf{p}_t := T(t)\mathbf{p}_0 = e^{-\Gamma(t)} \mathbf{p}_0 + [1-e^{-\Gamma(t)}]\, \mathbf{Q}(t) \ ,
\end{equation}
with
\begin{equation}\label{}
  \mathbf{Q}(t) = \frac{1}{1-e^{-\Gamma(t)}} \int_0^t \gamma(u) e^{\Gamma(u)} \mathbf{q}(u) du\ .
\end{equation}
and $\Gamma(t) = \int_0^t \gamma(u)du$. Equivalently one has
\begin{eqnarray}\label{TIQ}
  T(t) &=& e^{-\Gamma(t)} \mathbb{I}_2  + [1-e^{-\Gamma(t)}] \left( \begin{array}{rr} Q_1(t) & Q_1(t) \\ Q_2(t) & Q_2(t)    \end{array} \right)  \nonumber \\ & =& \left( \begin{array}{rr} Q_1(t) + e^{-\Gamma(t)} Q_2(t)  & Q_1(t) - e^{-\Gamma(t)} Q_1(t) \\ Q_2(t) - e^{-\Gamma(t)} Q_2(t) & Q_2(t) + e^{-\Gamma(t)} Q_1(t)    \end{array} \right)\ ,
\end{eqnarray}
where we have used
\begin{equation}\label{}
  {Q}_1(t) + Q_2(t) = 1\ .
\end{equation}
Formula (\ref{TIQ}) shows that $T(t)$ is a convex combination of two pseudo-stochastic matrices (actually, one of them $\mathbb{I}_2$ is stochastic). Now, $T(t)$ defines a legitimate dynamical map if and only if
\begin{equation}\label{CC}
  Q_1(t) + e^{-\Gamma(t)} Q_2(t) \geq 0 \ , \ \ \  Q_2(t) + e^{-\Gamma(t)} Q_1(t) \geq 0\ .
\end{equation}
for all $t \geq 0$. In particular, if
\begin{equation}\label{CCC}
  \Gamma(t) \geq 0\ , \ \  Q_1(t) \geq 0\ , \ \  Q_2(t) \geq 0\ ,
\end{equation}
then (\ref{CC}) is satisfied and $T(t) \in {\rm S}_2$.   It means that $\mathbf{Q}(t)$ is a legitimate state for all $t\geq 0$. Note, that if $x,y$ are time independent, then $\mathbf{q}(t) = \mathbf{Q}(t) = \mathbf{q}$ and
\begin{equation}\label{}
 \mathbf{p}_t = e^{-\gamma t} \mathbf{p}_0 + [1-e^{-\gamma t}]\, \mathbf{q} \ ,
\end{equation}
which means that the evolution is convex combination of the initial state $\mathbf{p}_0$ and the asymptotic invariant state $\mathbf{q}$ --- Markovian semigroup. Condition (\ref{CCC}) provides highly nontrivial constraints  for admissible functions $x(t)$ and $y(t)$. It is clear that if $x(t) \ngeq 0$ or $y(t) \ngeq 0$, then $T(t)$ is not divisible.
\end{ex}

Let $K \subset \Sigma_n$ be a convex set. We say that  a dynamical map $T(t)$ is $K$-divisible iff (\ref{TVT}) is satisfied with $V(t,s) \in {\rm PS}(K)$ for all $t \geq s$. Note, that if $K = \Sigma_n$ then $\Sigma_n$-divisibility reduces to divisibility. Moreover, if $K_1 \subset K_2$, then $K_2$-divisibility implies $K_1$-divisibility. Hence, if $T(t)$ is divisible then it is $K$-divisible for any $K$.

\begin{ex} Dynamical map $T(t)$ from Example 2 gives rise to the following family of propagators
\begin{equation}\label{}
  V(t,s) = \left( \begin{array}{rr} Q_1(t,s) + e^{-\Gamma(t,s)} Q_2(t,s)  & Q_1(t,s) - e^{-\Gamma(t,s)} Q_1(t,s) \\ Q_2(t,s) - e^{-\Gamma(t,s)} Q_2(t,s) & Q_2(t,s) + e^{-\Gamma(t,s)} Q_1(t,s)    \end{array} \right)\ ,
\end{equation}
where
\begin{equation}\label{}
  \Gamma(t,s) = \int_s^t \gamma(u)du\ , \ \ \ Q_k(t,s) = \frac{1}{1-e^{-\Gamma(t,s)}} \int_s^t \gamma(u) e^{\Gamma(u)} {q}_k(u) du\ .
\end{equation}
Taking $K_\varepsilon$ from Example 1 one finds that $K_\varepsilon$-divisibility provides extra constraints for $x(t)$ and $y(t)$ in order to $V(t,s) \in {\rm PS}(K_\varepsilon)$ for any $t \geq s$.
\end{ex}

\section{Pseudo-positive maps}

A linear map $\Phi : \mathfrak{B}(\mathcal{H}) \rightarrow \mathfrak{B}(\mathcal{H})$ is Hermitian if $(\Phi[A])^\dagger = \Phi[A^\dagger]$. It is positive iff $\Phi[A] \geq 0$ for all $A \geq 0$. Finally, it is trace-preserving if ${\rm tr}( \Phi[A]) = {\rm tr}A$. It is easy to show that positive maps are necessarily Hermitian. Note that $\Phi$ is PTP (Positive Trace-Preserving) iff for any orthonormal basis $\{e_1,\ldots,e_d\}$ in $\mathcal{H}$ the following matrix
\begin{equation}\label{T-Phi}
  T_{ij} := {\rm tr}(E_{ii} \Phi[E_{jj}]) \ ,
\end{equation}
is stochastic (we define the standard matrix units $E_{ij} := |e_i\>\<e_j|$). We call $\Phi$ pseudo-PTP if it is Hermitian and trace-preserving but not necessarily positive. Note, that $\Phi$ is pseudo-PTP iff the matrix $T_{ij}$ defined in (\ref{T-Phi}) is pseudo-stochastic. It is clear that pseudo-PTP maps form a semi-group, that is, if $\Phi_1$ and $\Phi_2$ are pseudo-PTP so is $\Phi_1 \Phi_2$.

Denote by $\mathfrak{S}$ a convex set of density operators in $\mathcal{H}$. It is clear that for any PTP map $\Phi$ maps $\mathfrak{S}$ into $\mathfrak{S}$. Now, let $K$ be a convex subset in $\mathfrak{S}(\mathcal{H})$ and let us define

\begin{enumerate}

\item $\mathcal{P}(K) \subset \mathcal{P}$ such that for all $\Phi \in \mathcal{P}(K)$ one has $\Phi[K] \subset K$,

\item $p\mathcal{P}(K) \subset p\mathcal{P}$ such that for all $\Phi \in p\mathcal{P}(K)$ one has $\Phi[K] \subset \mathfrak{S}$,

\item $\mathcal{P}_0(K) \subset \mathcal{P}(K)$ such that for all $\Phi \in \mathcal{P}_0(K)$ one has $\Phi[\mathfrak{S}] \subset K$,

\end{enumerate}
where $\mathcal{P}$ = PTP maps and $p\mathcal{P}$ = pseudo-PTP maps. Again, one has the following chain of inclusions

\begin{equation}\label{}
  \mathcal{P}_0(K) \subset \mathcal{P}(K) \subset \mathcal{P} \subset p\mathcal{P}(K) \subset p\mathcal{P}\ ,
\end{equation}
and if $K = \mathfrak{S}$, then
\begin{equation}\label{}
  \mathcal{P}_0(K) = \mathcal{P}(K) = \mathcal{P} \subset p\mathcal{P}(K) = p\mathcal{P}\ .
\end{equation}
If $K = \{ \rho_* = \frac 1d \mathbb{I}\}$ contains only maximally mixed state then $\mathcal{P}(K)$ defines a set of bistochastic positive maps and $\mathcal{P}_0(K)$ contains only one element $\Phi_*$ defined by
\begin{equation}\label{}
  \Phi_*[\rho] = \rho_* {\rm tr}\rho\ .
\end{equation}
Convex sets $\mathcal{P}_0(K)$ and $\mathcal{P}(K)$ contain only PTP maps. A set $p\mathcal{P}(K)$ contains also pseudo-PTP maps which are not positive. The interpretation of these maps is provided by the following

\begin{pro} A density operator $\rho \in \mathfrak{S}$ belongs to $K$ if and only if $\Phi[\rho] \in \mathfrak{S}$ for all $\Phi \in p\mathcal{P}(K)$.
\end{pro}

\noindent Hence, a map from $p\mathcal{P}(K) - \mathcal{P}$, i.e. pseudo-PTP but not positive,  may be used to witness that $\rho$ does not belong to $K$.

\begin{cor} A density operator $\rho \in \mathfrak{S}$ does not belong to $K$ if and only if there exists $\Phi  \in p\mathcal{P}(K) - \mathcal{P}$ such that $\Phi[\rho] \notin \mathfrak{S}$.
\end{cor}

\begin{ex} Consider $\mathcal{H} = \mathbb{C}^2$. In this case $\mathfrak{S}$ may be represented by the Bloch ball, that is,
\begin{equation}\label{}
  \rho = \frac 12 ( \mathbb{I} + \sum_{k=1}^3 x_k \sigma_k) \ ,
\end{equation}
and hence
\begin{equation}\label{}
  \mathfrak{S}  = \{ \, \mathbf{x} \in \mathbb{R}^3 \ | \ |\mathbf{x}|\leq 1\, \}\ .
\end{equation}
Now, consider a convex subset
\begin{equation}\label{}
  K_\varepsilon = \{ \, \mathbf{x} \in \mathbb{R}^3 \ | \ |\mathbf{x}|\leq 1-\varepsilon\, \} \subset \mathfrak{S}\ ,
\end{equation}
and let us analyze convex sets of pseudo-PTP bistochastic maps. Note, that density operators $\rho \in K_\varepsilon$ satisfy
$$  {\rm Purity}[\rho] = {\rm tr}\rho^2 \leq \frac{1+(1-\varepsilon)^2}{2} \ . $$
Equivalently, we may characterize this set via the von Neumann entropy: $\rho \in K_\varepsilon$ if
\begin{equation}\label{}
  S[\rho] \geq \ln 2 - \frac 12 [ (2-\varepsilon)\ln(2-\varepsilon) + \varepsilon\ln\varepsilon ]\ .
\end{equation}
A unital pseudo-PTP map $\Phi : \mathfrak{S} \rightarrow \mathfrak{S}$ may be represented in terms of the Bloch vectors as follows
\begin{equation}\label{}
  x'_k = \sum_{l=1}^3 A_{kl} x_l \ ,
\end{equation}
with $A_{kl}$ being matrix elements of $3 \times 3$ real matrix $A$. Now, Singular Value Decomposition of $A$ gives rise to
\begin{equation}\label{}
  A = O_1 D O_2^T\ ,
\end{equation}
where $O_1, O_2$ are orthogonal matrices and $D$ is the diagonal matrix of singular values $s_k$ of $A$. It is clear that $\mathbf{x}' \in K_\varepsilon$ iff the singular values $s_k$ satisfy
\begin{equation}\label{}
  s_k \leq \frac{1}{1-\varepsilon} \ ,
\end{equation}
for $k=1,2,3$.
\end{ex}

\begin{ex} Let us consider well known reduction map $\Phi : M_2(\mathbb{C}) \rightarrow M_2(\mathbb{C})$ defined by
\begin{equation}\label{}
  \Phi[\rho] = \mathbb{I}\, {\rm tr}\rho - \rho \ ,
\end{equation}
which is evidently positive since it maps any projector $|\psi\>\<\psi|$ to the orthogonal one $|\psi^\perp\>\<\psi^\perp| = \mathbb{I} - |\psi\>\<\psi|$. Let us define the family of trace-preserving maps
\begin{equation}\label{}
  \Phi_\mu[\rho] = \frac{1}{2-\mu} [\mathbb{I}\, {\rm tr}\rho - \mu\rho] \ ,
\end{equation}
for $\mu\in [1,2)$. Clearly these maps are pseudo-positive and only $\Phi_{\mu=1}$ is positive. One easily checks that for
\begin{equation}\label{mu}
 1 < \mu \leq \frac{1}{1+ (1-\varepsilon)^2}\ ,
\end{equation}
the map is not positive but $\Phi \in p\mathcal{P}(K_\varepsilon)$.  Hence, if $\mu$ satisfies (\ref{mu}) and $\Phi_\mu[\rho] \ngeq 0$, then $\rho \notin K_\varepsilon$, that is, the purity $P[\rho] > \frac{1}{2}[1+(1-\varepsilon)^2]$.

\end{ex}

\begin{remark} In the recent paper \cite{Adam} authors use the inverse to the reduction map in $M_n(\mathbb{C})$
\begin{equation}\label{}
  \Phi[X] = \frac{1}{n-1}[\mathbb{I}\, {\rm tr}X - X]  \ ,
\end{equation}
given by
\begin{equation}\label{}
  \Phi^{-1}[X] =  \mathbb{I}\, {\rm tr}X - (n-1)X \ ,
\end{equation}
to construct an entanglement witness in $\mathbb{C}^n \ot \mathbb{C}^n$. Note, that $\Phi^{-1}$ is not a positive map (unless $n=2$) but clearly it is pseudo-positive.
\end{remark}

\section{Non-Markovian $K$-divisible evolution}

Evolution of quantum system living in the Hilbert space $\mathcal{H}$ is described by the dynamical map, that is, a family of quantum channels
\begin{equation}\label{}
  \Lambda_t : \mathfrak{B}(\mathcal{H}) \to  \mathfrak{B}(\mathcal{H}) \ ,
\end{equation}
satisfying $\Lambda_0 = \oper$ (identity map). Consider now the dynamical map $\Lambda_t$ satisfying time-local master equation
\begin{equation}\label{}
  \dot{\Lambda}_t = L_t \Lambda_t \ ,
\end{equation}
with the time-dependent generator $L_t$. The map $\Lambda_t$ represents Markovian evolution if and only if $\Lambda_t$ is CP-divisible \cite{MARKOV1,MARKOV2,MARKOV3,MARKOV4} (see also \cite{rev1,rev2} for recent reviews), that is,
\begin{equation}\label{}
  \Lambda_t = V_{t,s} \Lambda_s \  ,
\end{equation}
and $V_{t,s}$ is completely positive for $t \geq s$. If the  maps $V_{t,s}$ are only positive then one calls $\Lambda_t$ P-divisible. Our approach enables one to generalize this notion: if $K$ is a convex subset of $\mathfrak{S}$, then one calls $\Lambda_t$ $K$-divisible iff $V_{t,s} \in p\mathcal{P}(K)$. If $K = \mathfrak{S}$, then $K$-divisibility reduces to P-divisibility. $K$-divisible evolution has the following property: $V_{t,s}$ maps any density operator from $K$ into the legitimate state. However, for $\rho \notin K$ the result of the action $V_{t,s}[\rho]$ needs not be a legitimate state.

\begin{ex} Consider the evolution of a qubit governed by the following generator

\begin{equation}\label{}
  L_t[\rho] = \frac 12 \sum_{k=1}^3 \gamma_k(t) [ \sigma_k \rho \sigma_k - \rho] \ ,
\end{equation}
with time dependent decoherence rates $\gamma_k(t)$. The corresponding solution reads \cite{PLA,PRA}
\begin{equation}\label{}
  \Lambda_t[\rho] = \sum_{\alpha=0}^3 p_\alpha(t) \sigma_\alpha \rho \sigma_\alpha\ ,
\end{equation}
with real $p_\alpha(t)$ and $\sum_{\alpha=0}^3 p_\alpha(t) =1$ given by
\begin{equation}\label{p-H}
  p_\alpha(t) =  \frac 14 \sum_{\beta=0}^3 H_{\alpha\beta} \lambda_\beta(t)\ ,
\end{equation}
and $H_{\alpha\beta}$ is the Hadamard matrix
$$	H=\left(\begin{array}{rrrr}1 & 1 & 1 & 1 \\1 & 1 & -1 & -1 \\1 & -1 & 1 & -1 \\1 & -1 & -1 & 1\end{array}\right)\ . $$
Finally, the quantities $\lambda_\alpha(t)$ define eigenvalues of the map $\Lambda_t$
\begin{equation}\label{}
  \Lambda_t[\sigma_\alpha]  =   \lambda_\alpha(t)\sigma_\alpha \ ,
\end{equation}
and they are given by: $\lambda_0=1$ (any trace-preserving Hermitian map satisfy this property) and
\begin{equation}\label{}
  \lambda_1(t) = \exp[- \Gamma_2(t,0) - \Gamma_3(t,0) ] \ \ +\ \mbox{cyclic perm.}
\end{equation}
with
\begin{equation}\label{}
  \Gamma_k(t,s) = \int_s^t \gamma_k(\tau) d\tau\ .
\end{equation}
One has the following result

\begin{enumerate}

\item $\Lambda_t$ is CP-divisible iff $\gamma_k(t) \geq 0$ for $k=1,2,3$,

\item $\Lambda_t$ is P-divisible iff

$$\gamma_1(t) +\gamma_2(t) \geq 0\ , \ \ \gamma_2(t) +\gamma_3(t) \geq 0\ , \ \ \gamma_3(t) +\gamma_1(t) \geq 0\ , $$

\item $\Lambda_t$ is $K_\varepsilon$-divisible iff

$$   \Gamma_1(t,s) + \Gamma_2(t,s) \geq \ln[1-\varepsilon] \ , \ \Gamma_2(t,s)  + \Gamma_3(t,s) \geq \ln[1-\varepsilon]  \ , \ \Gamma_3(t,s) + \Gamma_1(t,s) \geq \ln[1-\varepsilon] \  , $$ 
for $t > s $.

\end{enumerate}
It is clear that if $\varepsilon \to 0$, then 3. reduces to 2. Conversely, if  $\varepsilon \to 1$, then $\gamma_k(t)$ are completely arbitrary.

\end{ex}

\section{Conclusions}

Stochastic matrices preserve  the simplex of probability vectors in $\mathbb{R}^n$. Similarly, trace-preserving positive maps preserve the convex set of density matrices in in $\mathfrak{B}(\mathcal{H})$. It is therefore clear that both objects proved to be very important for the analysis of various properties of classical and quantum systems. In this paper we introduced the notions of pseudo-stochastic $n \times n$ matrices and pseudo-positive maps acting in $\mathfrak{B}(\mathcal{H})$. These objects provide a natural generalization of stochastic matrices and positive maps. They naturally appear
when we only consider the transformation of a convex subdomains in the set of states. Actually, one is often interested not in the whole set of states but only in a suitable convex subset satisfying some extra properties (like for example additional symmetries and/or special preparation procedure). In a realistic laboratory scenario one usually has an access only to a subset of states defined by the admissible preparation scheme. Therefore, it is natural to extend the notion of stochastic matrices and positive maps to deal with more general scenarios as well.  Interestingly, these more general matrices or maps may be used as witnesses that a given state does not belong to a convex subdomain in perfect analogy to entanglement witnesses.

Moreover, given a dynamical map --- classical $T(t)$ or quantum $\Lambda(t)$ --- the corresponding propagators $T(t,s)=T(t) T^{-1}(s)$ and  $\Lambda(t,s)=\Lambda(t) \Lambda^{-1}(s)$ are always pseudo-stochastic and pseudo-positive, respectively. We have shown that these objects are useful for the refinement of the notion of divisible maps and hence may be used to further characterization of non-Markovian classical and/or quantum evolution.
Indeed, if $T(t,s)$ is stochastic for any $t>s$, then classical evolution is Markovian. Similarly, if $\Lambda(t,s)$ is positive, then quantum evolution is P-divisible which is considered as a natural notion of Markovianity in the quantum case   \cite{rev2}.  Possible new applications are currently investigated.

\section*{Acknowledgements}

DC was partially supported by the National Science Center project
DEC-2011/03/B/ST2/00136. V.M. thanks University Federico II in Naples and INFN for hospitality.
We thank Dorota Chru\'sci\'nska for preparing the figure.

\section*{Appendix. Lie algebra of the "diamond" group}

In this Appendix we present the structure of pseudo-stochastic matrices for $n=2$ and $n=3$. For the qubit case one has the matrix $T_2$ of the form
  \begin{eqnarray}
  T_2=\left(
\begin{array}{cc}
 1-a & b \\
 a & 1-b \\
\end{array}
\right).
\end{eqnarray}
One can measure the pseudo-stochasticity by means of the negativity, $-\max (|a|, |1-a|)$. If the matrices are written for the Lie group, the generators of the Lie algebra and their commutator read
 \begin{eqnarray}
  L_a=\left(
\begin{array}{cc}
 -1 & 0 \\
 1 & 0 \\
\end{array}
\right), \ \
 L_b=\left(
\begin{array}{cc}
 0& 1 \\
 0& -1 \\
\end{array}
\right), \ \ [L_a,L_b]=L_a-L_b.
\end{eqnarray}
This is a  solvable Lie algebra corresponding to the Lie algebra of the solvable group of matrices
 \begin{eqnarray}
  g=\left(
\begin{array}{cc}
 a & b \\
 0 & 1 \\
\end{array}
\right).
\end{eqnarray}
Another example is the qutrit case. Then, the pseudo-stochastic matrix has the form
\begin{eqnarray}
T_3=\left(
\begin{array}{ccc}
 1-a_1-a_2 & b_1 & c_1 \\
 a_1 &1 -b_1-b_2 & c_2 \\
 a_2 & b_2 & 1-c_1-c_2 \\
\end{array}
\right).
\end{eqnarray}
The six generators of the Lie algebra read
\begin{eqnarray}
{L_1}=\left(
\begin{array}{ccc}
 -1 & 0 & 0 \\
 1 & 0 & 0 \\
 0 & 0 & 0 \\
\end{array}
\right), \
{L_2}=\left(
\begin{array}{ccc}
 -1 & 0 & 0 \\
 0 & 0 & 0 \\
 1 & 0 & 0 \\
\end{array}
\right), \
{L_3}=\left(
\begin{array}{ccc}
 0 & 1 & 0 \\
 0 & -1 & 0 \\
 0 & 0 & 0 \\
\end{array}
\right),\\
{L_4}=\left(
\begin{array}{ccc}
 0 & 0 & 0 \\
 0 & -1 & 0 \\
 0 & 1 & 0 \\
\end{array}
\right), \
{L_5}=\left(
\begin{array}{ccc}
 0 & 0 & 1 \\
 0 & 0 & 0 \\
 0 & 0 & -1 \\
\end{array}
\right),\
{L_6}=\left(
\begin{array}{ccc}
 0 & 0 & 0 \\
 0 & 0 & 1 \\
 0 & 0 & -1 \\
\end{array}
\right). \notag
\end{eqnarray}
One easily finds for the commutation relations:
\begin{eqnarray}
&& \left[L_1, L_2\right]= L_2-L_1,[L_1, L_3]= L_1-L_3, [L_1,L_4]= L_1-L_2, [L_1,L_5]= L_6-L_5, \\ &&\left[L_1,L_6\right]= 0,
 \left[L_2, L_3\right]= L_4-L_3, [L_2, L_4]= 0, [L_2,L_5]= L_2-L_5, [L_2,L_6]= L_2-L_1, \notag\\
&& \left[L_3, L_4\right]= L_4-L_3, [L_3, L_5]= 0, [L_3,L_6]= L_5-L_6,
 \left[L_4, L_5\right]= L_4-L_3, [L_4,L_6]= L_4-L_6, \notag\\
&& \left[L_5, L_6\right]= L_6-L_5, \notag
\end{eqnarray}
One can see that the Lie algebra has several three--dimensional subalgebras, for example those given by three generators corresponding to the subgroup of elements
\begin{eqnarray}
\tilde{T}_3=\left(
\begin{array}{ccc}
 1& b_1 & c_1 \\
 0&1 -b_1& c_2 \\
0& 0& 1-c_1-c_2 \\
\end{array}
\right).
\end{eqnarray}
There are two-dimensional solvable subalgebras corresponding,  for example, to the subgroup
\begin{eqnarray}
\hat{T} =\left(
\begin{array}{ccc}
 1& 0 & c_1 \\
 0&1 & c_2 \\
0& 0& 1-c_1-c_2 \\
\end{array}
\right) ,
\end{eqnarray}
along with two--dimensional subalgebras corresponding to the bistochastic matrices
\begin{eqnarray}
T_B=\left(
\begin{array}{ccc}
 1-a_1-a_2 & a_2 & a_1 \\
 a_1 &1 -a_1-a_2 & a_2 \\
 a_2 & a_1 & 1-a_1-a_2 \\
\end{array}
\right).
\end{eqnarray}
In the case of qudits, analogous properties of the Lie algebra can be established following, for instance, our discussion of stochastic matrices embedding into the affine group \cite{semigroup}. One can extend the present approach to the case of an infinite--dimensional simplex. For instance, we can include the analogous description in the framework of the tomographic picture of the Gaussian or other quantum states and pseudo-positive maps relating the states.

\end{document}